\documentclass[letterpaper, 10 pt, conference]{ieeeconf}  % Comment this line out if you need a4paper

\IEEEoverridecommandlockouts                              % This command is only needed if 
\title{\LARGE \bf
Evaluation of Connected Vehicle Identification-Aware Mixed Traffic Freeway Cooperative Merging
}

\author{Haoji Liu$^{1}$, Fatemeh Jahedinia$^{2}$, Zeyu Mu$^{1}$, and B. Brian Park$^{1}$$^{2}$$^{*}$  % <-this % stops a space
\thanks{This work was supported by the U.S. National Science Foundation under Grant No. CMMI-2009342.}% <-this % stops a space
\thanks{$^{1}$H. Liu, Z. Mu, and B. B. Park are with the Link Lab and the Department of Systems and Information Engineering, University of Virginia, Charlottesville, VA 22903 USA.} 
\thanks{$^{2}$F. Jahedinia and B. B. Park are with the Department of Civil and Environmental Engineering, University of Virginia, Charlottesville, VA 22903 USA. (e-mail: 
        {\tt\small haojiliu@email.virginia.edu, bfg9xt@virginia.edu, dwe4dt@virginia.edu, bp6v@virginia.edu} ).
        $^{*}$Corresponding Author: B. Brian Park.}}%

\begin{document}
\maketitle
% *** IEEE Copyright notice with TikZ ***
% 
\copyrightnotice
\thispagestyle{empty}
\pagestyle{empty}
%%%%%%%%%%%%%%%%%%%%%%%%%%%%%%%%%%%%%%%%%%%%%%%%%%%%%%%%%%%%%%%%%%%%%%%%%%%%%%%%
\begin{abstract}

Cooperative on-ramp merging control for connected automated vehicles (CAVs) has been extensively investigated. However, they did neglect the connected vehicle identification process, which is a must for CAV cooperations. In this paper, we introduced a connected vehicle identification system (VIS) into the on-ramp merging control process for the first time and proposed an evaluation framework to assess the impacts of VIS on on-ramp merging performance. First, the mixed-traffic cooperative merging problem was formulated. Then, a real-world merging trajectory dataset was processed to generate dangerous merging scenarios. Aiming at resolving the potential collision risks in mixed traffic where CAVs and traditional human-driven vehicles (THVs) coexist, we proposed on-ramp merging strategies for CAVs in different mixed traffic situations considering the connected vehicle identification process. The performances were evaluated via simulations. Results indicated that while safety was assured for all cases with CAVs, the cases with VIS had delayed initiation of cooperation, limiting the range of cooperative merging and leading to increased fuel consumption and acceleration variations. 

\end{abstract}


%%%%%%%%%%%%%%%%%%%%%%%%%%%%%%%%%%%%%%%%%%%%%%%%%%%%%%%%%%%%%%%%%%%%%%%%%%%%%%%%
\section{INTRODUCTION}

The connected automated vehicle (CAV) is a revolutionary technology for shaping our future transportation. Apart from CAV’s capabilities of automatically perceiving driving environments, planning trajectories, and controlling motions, it also yields benefits through its vehicle-to-everything (V2X) connectivity, which can obtain beyond-vision-range traffic information, thus making decisions as early as possible to avoid any potential risks. Furthermore, with the cooperation of multiple CAVs, it is possible to improve traffic mobility and the environment in a wide range \cite{guanetti2018control}.

A typical scenario of CAV cooperation is the freeway merging. In this scenario, vehicles on the freeway mainline and on-ramp compete for passing priority, thereby often resulting in unsafe conditions \cite{liu2023safety}. Given the on-ramps deviate from the adjacent mainline and many physical obstacles such as trees or advertisement signs block the line of sight, the freeway merging has potential risks of collision. However, with the V2X communications, CAVs can share basic safety messages (BSM) including motion data with each other, so the situation of inadequate traffic information can be effectively avoided, leading to mitigation of traffic conflicts at on-ramp merging areas.

Existing studies on CAV on-ramp merging cooperation cover full-CAV scenarios and mixed traffic scenarios where CAVs and traditional human-driven vehicles (THVs) coexist. The research topic mainly includes sequence and trajectory planning. Sequence planning determines the order that vehicles from main roads and on-ramps pass through the merging area \cite{rios2016survey}. By specifying a sequence, merging conflicts could be resolved, and travel efficiency could be improved. Typical sequence planning methods include rule-based \cite{ding2019rule}, optimization-based \cite{pei2019cooperative}, game theory-based \cite{jing2019cooperative}, and graph-based methods \cite{shi2023cooperative}. Based on a merging sequence, merging trajectories for CAVs can be planned based on optimal control \cite{xiao2021decentralized}, model predictive control \cite{liu2018strategy}, control barrier functions \cite{liu2021decentralized}, or reinforcement learning \cite{chen2023deep}, to realize safe, energy efficient, and comfortable driving.

\begin{figure}
\centering
\includegraphics[width=3.4in]{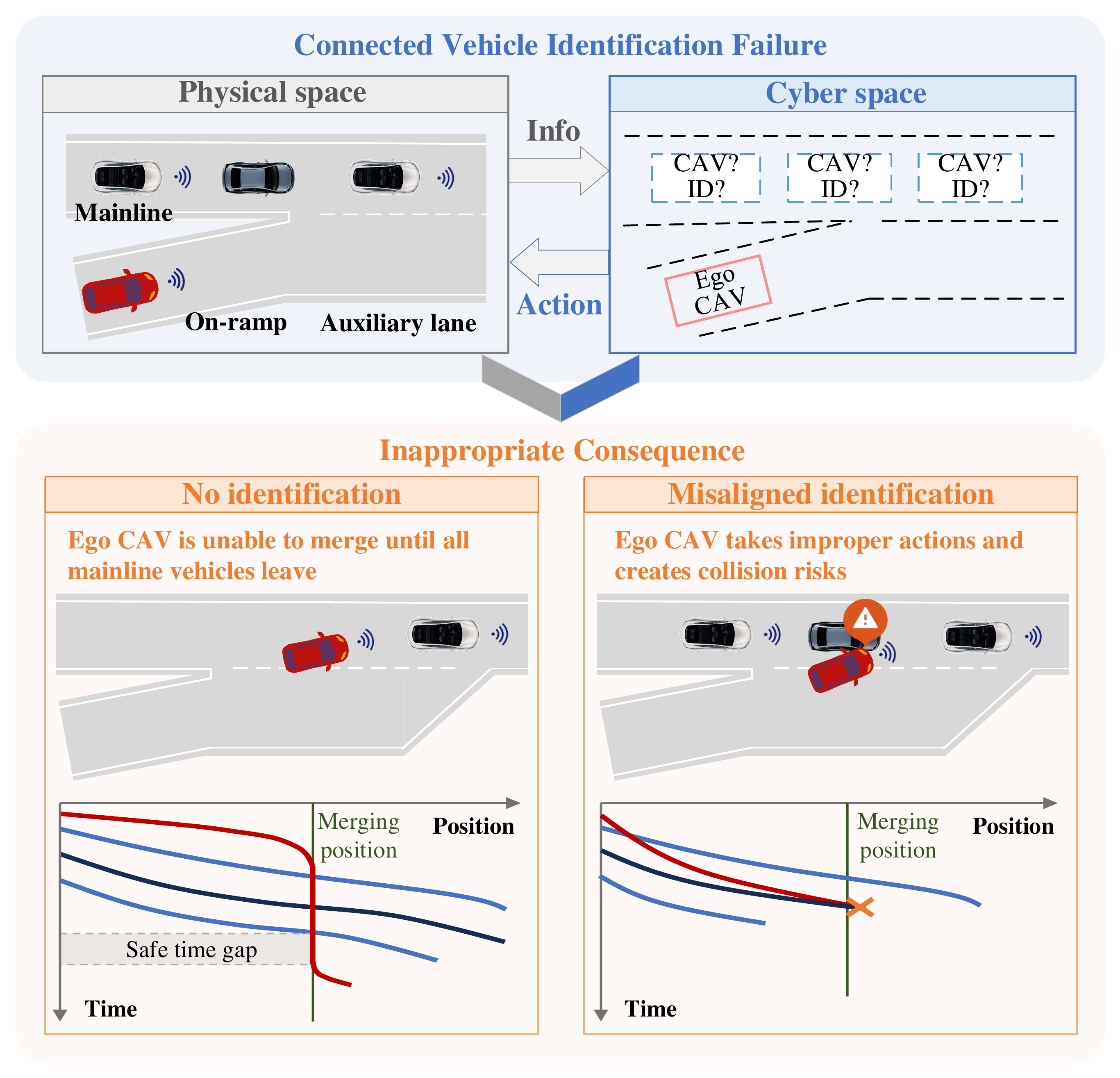}
\captionsetup{justification=justified}
\caption{An example of a vehicle identification failure in the on-ramp merging area and its negative impact. The ego on-ramp CAV fails to identify which vehicle is a CAV and which CAV broadcasts which message, therefore possible bad consequences may happen. }\label{fig1_Example}
\end{figure}

Although existing research revealed the positive impact of cooperative CAV on-ramp merging control on travel safety, travel efficiency, and energy efficiency improvement, a significant prerequisite was neglected – \textit{how to properly identify which CAV to cooperate with}. That is, an ego vehicle needs to know received connected vehicle information is coming from which CAV. This seemingly trivial issue is, in fact, of paramount importance. An example is illustrated in Fig. \ref{fig1_Example}. In physical space, all mainline CAVs simultaneously broadcast messages to the ego on-ramp CAV through V2V communications. However, without vehicle identification capability, the ego CAV cannot accurately identify which vehicle is a CAV, and which CAV broadcasts which message in cyberspace. Consequently, the on-ramp CAV cannot cooperatively merge. The other consequence is an ego CAV on-ramp merging vehicle incorrectly identifies the mainline vehicles and results in a potential collision with mainline vehicles. Therefore, precise CAV identification is of utmost importance for merging control.

A straightforward method to achieve CAV identification is comparing the data perceived by on-board sensors of the ego vehicle and BSM received by the ego CAV through V2V communications. When there is no measurement error, surrounding CAVs can be accurately identified through directly matching the positions calculated based on radar and V2V-BSM information. However, the reality is that although the radar detection error is small (around 0.1m), the regular GPS positioning error is approximately $1\sim5$ m \cite{rychlicki2020analysis}. So, it is almost impossible to identify vehicles by simply matching these two kinds of data. In addition, in mixed traffic scenarios, THVs will cause interference in the identification of CAVs. Therefore, how to identify CAVs quickly and accurately is a challenge.

In our previous research \cite{chen2022connected}, a preceding vehicle identification system (VIS) was proposed to address the challenge of preceding CAV identification for platooning in highway mixed-traffic environments. Its principle is that, based on the surrounding vehicles’ data obtained via radar perception and V2V communication, an ego vehicle will identify the V2V data source of the preceding vehicle and remember its unique message ID. After the completion of vehicle identification, the ego vehicle can utilize preceding vehicle’s motion information to achieve cooperative adaptive cruise control (CACC).

However, compared to the CACC application, considering VIS will bring more troubles to the cooperative on-ramp merging task. To be specific, when implementing cooperative platooning tasks in freeways, the line of sight is adequate for radar to detect surroundings, and the road length is long enough to execute cooperation strategies after vehicle identification. For the cooperative merging task, on the contrary, radar perception can only work after on-ramp vehicles obtaining adequate line of sight due to possible obstacles between the main road and on-ramp. This means the VIS may only identify on-ramp vehicles in a short range, resulting in a short time for CAV cooperation after identification. This constraint raises the potential risks associated with the cooperative merging process. Hence, it's crucial to conduct a comprehensive assessment of the VIS's impact, particularly in hazardous scenarios, and to develop appropriate control strategies to safeguard cooperative merging in mixed traffic. Furthermore, while there are evaluation frameworks for hardware deployment \cite{wang2022mobility}, software co-simulation \cite{chen2024safety}, and performance measurement \cite{tian2018performance} in cooperative merging scenarios, there is still a need for an evaluation framework specifically focusing on connected vehicle identification-aware cooperative merging in dangerous scenarios within mixed traffic environments.

To address the identified research gaps, this paper introduces an evaluation framework that encompasses dangerous merging scenario generation, classification of mixed traffic merging cases, integration of operational strategies, and evaluation metrics. Based on this framework, we analyzed the impact of the VIS on the CAV merging process. Specifically, we discussed the possible control strategies for CAVs considering VIS implementation in different mixed-traffic environments. A real-world Exits and Entries Drone (exiD) dataset \cite{moers2022exid}, which includes a bunch of naturalistic on-ramp merging trajectory data, was used to evaluate the proposed merging strategies. This research would help bridge the gap between the theoretical investigation and real-world application of cooperative on-ramp merging control of CAVs.

The contributions of this paper are twofold. 1) To the best of our knowledge, the vehicle identification system (VIS) was introduced into the cooperative on-ramp merging process for the first time. We elucidated the challenges posed by the application of VIS in on-ramp scenarios for merging control and described the VIS algorithm applied in mixed traffic. Based on this, we proposed CAV merging control strategies for various cases in mixed traffic considering the impact of vehicle identification time. 2) An evaluation framework is proposed to assess the impact of implementing VIS on the mixed traffic cooperative merging task. We first abstracted dangerous merging scenarios from the real-world merging trajectory dataset, serving as the baseline for evaluation. Subsequently, our proposed merging operation strategies were verified to show their effectiveness in eliminating merging safety hazards. The results uncovered the potential benefits of VIS, and we also discussed approaches to facilitate these benefits. This discussion could offer valuable guidance for future research in the field of cooperative merging.

The remainder of this paper is organized as follows. In Section \ref{section2}, the on-ramp merging problem in mixed traffic is formulated and a framework of merging performance evaluation is proposed. In Section \ref{section3}, the naturalistic trajectory data is processed to generate dangerous merging scenarios. The operation strategies for CAVs in various mixed-traffic environments are designed, and the VIS algorithm for on-ramp merging is described. Simulations are conducted and analyzed in Section \ref{section4}. Finally, conclusions are made in Section \ref{section5}.

\section{PROBLEM DESCRIPTION} \label{section2}

In this section, the on-ramp merging scenario in mixed traffic is described. Then, vehicle dynamics and control modes for CAVs in different mixed traffic conditions are illustrated. After that, an evaluation framework for vehicle identification-aware CAV merging control strategies is proposed.

\subsection{Scenario description}

\begin{figure}
\centering
\includegraphics[width=3in, trim=0cm 2cm 0cm 2cm, clip]{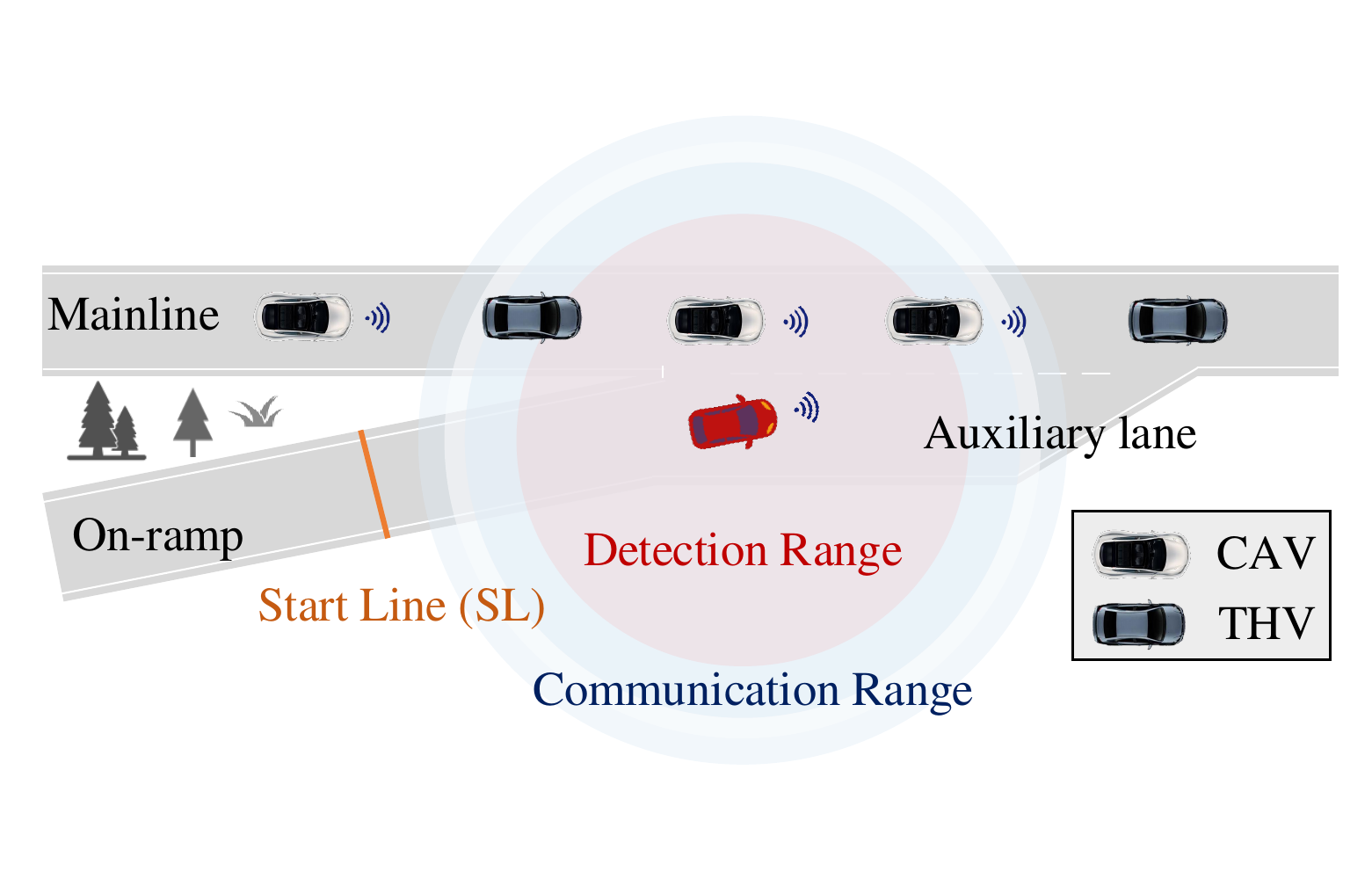}
\captionsetup{justification=justified}
\caption{The on-ramp merging scenario. From the Start Line, on-ramp CAVs can detect nearby mainline vehicles with good line of sight.}\label{fig2_scenario}
\end{figure}

In this paper, we consider a common on-ramp merging scenario with one mainline and one on-ramp connecting with an auxiliary lane \cite{zhou2018optimal}, as shown in Fig. \ref{fig2_scenario}. In the mixed traffic environment, where both CAVs and THVs coexist, CAVs can share BSM with other CAVs through V2V communication. On the contrary, THVs do not have wireless communication capability. Additionally, CAVs can perceive surrounding vehicles via radar detection only when the line of sight is adequate. In this context, the focus of this paper is on the issue of an on-ramp CAV attempting to merge into the mainline.

To implement VIS by utilizing V2V communication and radar detection, physical requirements should be mentioned. The communication range of a CAV could be several hundred meters, which means on-ramp CAVs can receive V2V message of mainline CAVs at the beginning of the on-ramp. However, the radar detection range is relatively small and is only effective when the line of sight is adequate. Therefore, a Start Line (SL) is marked in Fig. \ref{fig2_scenario}, from which the line of sight is good enough so that on-ramp CAVs can detect nearby mainline vehicles and start identifying them via VIS.

\subsection{Vehicle dynamics and merging constraints of CAVs} 

\noindent (1) Vehicle dynamics

Since we need to design a controller for CAVs, vehicle dynamics and merging constraints for CAVs are modelled here. As our emphasis is on longitudinal movement adjustment, a 2$^{\text{nd}}$-order linear longitudinal vehicle model is used for CAV controller design \cite{rios2016survey}, which is represented as:
\begin{equation} \label{eq:1}
\begin{aligned}
\dot{X}=\left[\begin{array}{l}
\dot{p}(t) \\
\dot{v}(t)
\end{array}\right]=\left[\begin{array}{l}
v(t) \\
u(t)
\end{array}\right]
\end{aligned}
\end{equation}
where $p(t)$ and $v(t)$ are the position and speed of the CAV, respectively. $u(t)$ is the acceleration of the CAV, which is also regarded as the control input of the CAV system.

\noindent (2) Safety constraints

At the moment of the on-ramp vehicle merging into the mainline, there should be a safe inter-vehicle distance between the adjacent mainline vehicle and on-ramp vehicle. Since we want to facilitate a seamless transition of the on-ramp CAVs into the car-following mode after completing the merging control maneuver, we set the terminal merging speed of the on-ramp vehicle $v_r(t_{mer})$ as same as that of its nearby mainline vehicle $v_m(t_{mer})$ and apply the time gap policy \cite{vogel2003comparison}, which is widely used in merging control and car-following control studies. In this way, the terminal speed constraint is
\begin{equation} \label{eq:2}
\begin{aligned}
v_r (t_{mer} )= v_m (t_{mer})
\end{aligned}
\end{equation}
\noindent where $t_{mer}$ is the time when the merging CAV merges into the mainline (also called merging time). Based on time gap policy, if an on-ramp vehicle is behind a mainline vehicle at the merging time, the distance between these two vehicles satisfies:
\begin{equation} \label{eq:3}
\begin{aligned}
\frac{p_{m}\left(t_{m e r}\right)-p_{r}\left(t_{m e r}\right)-l}{v_{r}\left(t_{m e r}\right)} \geq h
\end{aligned}
\end{equation}
\noindent where  $p_m(t_{mer})$ and $p_r(t_{mer})$ are the position of the mainline preceding vehicle and the on-ramp following vehicle, respectively, $l$ is the vehicle length, set at 2.5 meters. $h$ is the safe merging time gap, set at 1.8 seconds \cite{zhou2018optimal}.

If an on-ramp vehicle is ahead of a mainline vehicle at the merging time, the inter-vehicle distance should satisfy:
\begin{equation} \label{eq:4}
\begin{aligned}
\frac{p_{r}\left(t_{m e r}\right)-p_{m}\left(t_{m e r}\right)-l}{v_{m}\left(t_{m e r}\right)} \geq h
\end{aligned}
\end{equation}
\subsection{Control modes for CAVs}

Different control modes should be designed to match various mixed-automated traffic conditions. This paper focuses on two types of vehicles with different automation levels: CAVs and THVs. Focused on providing a safe and efficient merging trajectory to the on-ramp CAV, two types of control modes are proposed: recursive control and cooperative control.

\noindent (1) Control Mode I: Recursive Control.

The recursive control mode is applied to scenarios where the on-ramp CAV needs to merge near a mainline THV. Since the movement of the THV is uncertain, the on-ramp CAV has to repeatedly sample the states of the THV, based on which its control inputs can be calculated recursively. 

Denote the recursive time step as $\Delta t$. Each control interval is denoted as ($t_s$,$t_s + \Delta t$], where the $t_s=t_0+k \Delta t (k=0,1,2,…)$ is the start time of this interval, and $t_0$ is the initial time for merging control. Assume that within each control interval, the mainline THV maintains the speed at $t_s$, the optimal control law is applied to calculate the optimal control input of the on-ramp CAV. To deal with uncertain movements of THVs, only the first control input is adopted for this control step. After that, the controller jumps into the next control interval, and the new control input is calculated based on the updated states of mainline THVs. Such a recursive process is repeated until the on-ramp vehicle and its surrounding mainline THV have proper position and speed differences. With the goal of minimizing energy consumption, the mathematical problem of Control Mode I is given \cite{xiao2021decentralized}.

\textit{Control Problem I. Recursive merging control}
\begin{equation} \label{eq:5}
\begin{aligned}
\underset{u_{r}}{\operatorname{\argmin}} \sum_{t_{s}=t_{0}+k \Delta t}^{t_{m e r}}\left(\frac{1}{2} u_{r}^{2}\right), k=0,1,2, \ldots
\end{aligned}
\end{equation}

\noindent s.t.

\noindent \hspace{0.1em} Vehicle dynamics: Eq.(\ref{eq:1})

\noindent \hspace{0.1em} Initial states of a control step: $p_r (t_s )$,$v_r (t_s )$

\noindent \hspace{0.1em} Terminal speed constraint: Eq.(\ref{eq:2})

\noindent \hspace{0.1em} Terminal position constraint:

\noindent \hspace{0.1em} - If the on-ramp CAV merges after a mainline vehicle: Eq.(\ref{eq:3})

\noindent \hspace{0.1em} - If a mainline vehicle merges after the on-ramp CAV: Eq.(\ref{eq:4})

Note that to meet the terminal speed constraint, the terminal speed of the on-ramp CAV is reset in every control step based on the speed of the mainline vehicle.

\noindent (2) Control Mode II: Cooperative Control. 

If the on-ramp CAV needs to merge near a mainline CAV, the cooperative control mode may be applied, contingent upon whether these two CAVs are equipped with VIS. (a) If neither of the two CAVs has VIS, it implies uncertainty regarding the origin of V2V messages. Considering the localization errors in GPS, blindly trusting the vehicle positioning information provided by V2V-BSM for cooperative control is highly risky. Therefore, the control mode for the on-ramp CAV should be Control Mode I: recursive control. (b) If all CAVs are equipped with VIS, these two CAVs can cooperate after completing vehicle identifications.

In Control Mode II, when the on-ramp CAV identifies its surrounding mainline vehicle as a CAV and builds connection with it, cooperative control can be applied to realize energy-efficient merging. The merging time $t_{mer}$ and position $p_r (t_{mer} )$ can be preset based on traffic conditions.

\textit{Control Problem II. Cooperative merging control}
\begin{equation} \label{eq:6}
\begin{aligned}
\underset{u_{r}}{\operatorname{argmin}} \int_{t_{0}}^{t_{m e r}}\left(\frac{1}{2} u_{r}^{2}\right) d t
\end{aligned}
\end{equation}
\noindent s.t.

\noindent \hspace{0.1em} Vehicle dynamics: Eq.(\ref{eq:1})

\noindent \hspace{0.1em}  Initial states of a control step: $p_r (t_0 )$,$v_r (t_0 )$

\noindent \hspace{0.1em} Terminal speed constraint: Eq.(\ref{eq:2})

\noindent \hspace{0.1em} Terminal position constraint:

\noindent \hspace{0.1em} - If the on-ramp CAV merges after a mainline vehicle: Eq.(\ref{eq:3})

\noindent \hspace{0.1em} - If a mainline vehicle merges after the on-ramp CAV: Eq.(\ref{eq:4})

Note that compared to Control Problem I which is a recursive optimal control problem solved at each start time $t_s=t_0+\Delta t$, Control Problem II only involves one-time calculation at the initial time for merging control $t_0$. Both two problems can be analytically solved by using Pontryagin’s maximum principle (PMP), which is computationally efficient. The readers can refer to \cite{xiao2021decentralized} for more details. $\Delta t$ is set 1/25 second in this paper, which is the same as the sampling rate of the exiD trajectories.

\subsection{Evaluation Framework}

Recall that in this paper, our primary focus is to evaluate the impact of the connected vehicle identification process on merging performance in mixed traffic scenarios. Specifically, we aim to propose distinct operational strategies for CAVs, taking into account various mixed traffic scenarios and the presence or absence of the VIS. Building upon these strategies, we conduct a fair comparison with real-world merging data to validate performance differences in safety, energy efficiency, and comfort among the proposed operational strategies and across diverse traffic scenarios. Through the results, our goal is to uncover the benefits of VIS for CAV cooperation and explore the challenges it introduces to the cooperative merging control of CAVs, considering the influence of the VIS. 

\begin{figure}
\centering
\includegraphics[width=3.5in]{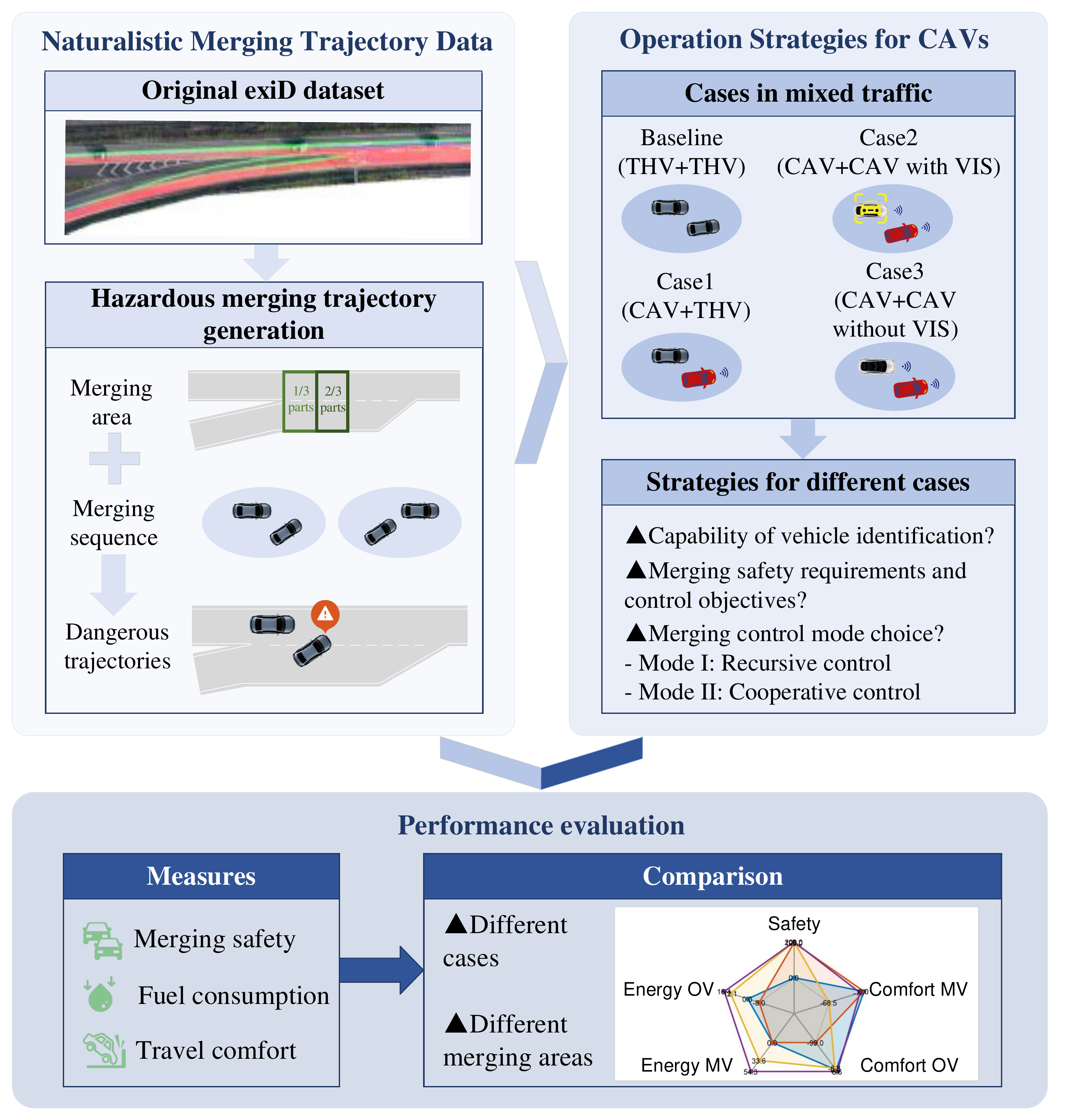}
\caption{Evaluation framework. 
}\label{fig3_framework}
\end{figure}

To achieve these objectives, our methodology follows the evaluation framework depicted in Fig.~\ref{fig3_framework}. This process encompasses merging trajectory data filtering and processing, design of evaluation cases, formulation of operational strategies, and ultimately, the final evaluation.

\section{DATA AND METHOD} \label{section3}
In this section, baseline merging trajectories are generated based on real-world merging data. Operation strategies for on-ramp CAVs in different mixed-traffic environments are proposed, and the vehicle identification system (VIS) is described. 

\subsection{Real-World Merging Data}

\noindent (1) Data and scenario setting

The real-world trajectory data we explored is the exiD dataset, which is a real-world trajectory dataset of highly interactive highway scenarios in Germany \cite{moers2022exid}. It contains trajectories of 1,113 vehicles and spans over 65,000 seconds of driving time. The trajectories in the dataset are highly interactive, with frequent lane changes, merging, and other complex maneuvers. 

According to our observations of the exiD dataset, there are three primary merging areas for the on-ramp vehicles, which can be seen as different driving preferences of drivers \cite{mu2023does}. That is, some may aggressively merge around the start point of the auxiliary lane; some may moderately observe the condition of the mainline and then choose to merge into the mainline at the middle of the merging zone; and the others may conservatively yield to the mainline vehicles and wait a safe enough chance for merging, which may lead to late merging at the end of the auxiliary lane. 

\begin{figure}
\centering
\includegraphics[width=3.4in]{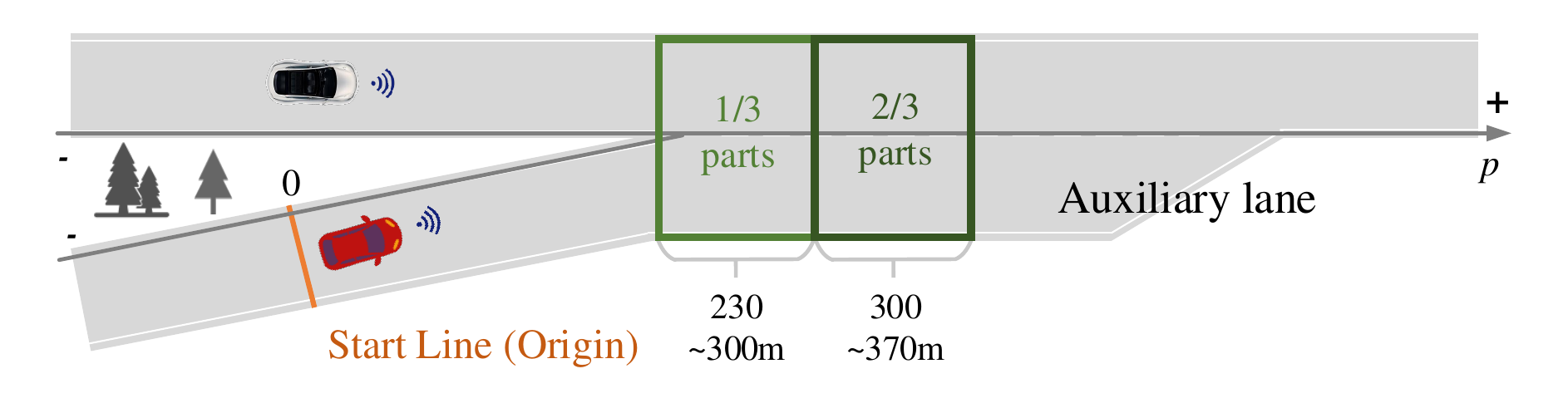}
\caption{Two merging parts in the auxiliary lane.  
}\label{fig4_twoparts}
\end{figure}

Based on the three kinds of driver aggressiveness preferences, we can roughly divide the auxiliary lane into three sections. Set the Start Line as origin, i.e., $p = 0$. Because the one-third part ($p = 230\sim300$ meters) and two-third part ($p = 300\sim370$ meters) correspond to the aggressive and moderate merging preferences, which has a higher risk of collision, we mainly simulate the on-ramp vehicle has the merging position within these two parts (as shown in Fig. \ref{fig4_twoparts}) and evaluate how these merging preferences affect the safety and other merging performance of the two vehicles.

\noindent (2) Baseline trajectory generation

To evaluate the effectiveness of merging control strategies in mitigating safety hazards during merging, a task of utmost importance in merging control, we established a baseline by generating collision-risk merging trajectories based on the exiD dataset. To focus on instances where both the mainline vehicle and the on-ramp vehicle exhibit collision risks at the merging area, we compiled a data set that includes 100 pairs of vehicles in one-third of the auxiliary lane and another 100 pairs of vehicles in two-thirds of the auxiliary lane. Among these pairs, 18\% involve the on-ramp CAV merging after its potentially conflicting mainline vehicle, while 82\% depict scenarios where the on-ramp CAV merges ahead of its potentially conflicting mainline vehicle. 

\subsection{Operation Strategies for on-ramp CAVs}

As the on-ramp CAV encounters potential collision risks with mainline vehicles, appropriate operational strategies need to be devised based on the automation level of the mainline vehicle and the CAV's capability in vehicle identification. Generally, there are 3 cases to be considered.

\noindent (1) Case 1: CAV-THV

In this case, the on-ramp CAV might have collision risks with a mainline THV. However, since the on-ramp CAV may receive V2V-BSM from other mainline vehicles, it still needs to operate VIS to check whether this nearest mainline vehicle is a THV or a CAV. After identifying this mainline vehicle is a THV, the on-ramp CAV will adopt Control Mode I to recursively generate merging trajectory.

\noindent (2) Case 2: CAV-CAV with VIS

In Case 2, the on-ramp CAV is exposed to potential collision risks with a mainline CAV. The on-ramp CAV should first identify which V2V-BSM is sent by this mainline CAV. After that, they can build V2V directed connection and plan their future motion cooperatively via Control Mode II. 

\textit{Remark:} There are two situations that classify a CAV-CAV pair as Case 1. The first situation occurs when at least one of the paired CAVs experiences from V2V communication failure and degrades to an AV. Consequently, the on-ramp CAV is required to recursively detect the states of the mainline CAV and plan its merging trajectory. The second situation arises when two CAVs can communicate effectively, but the VIS of both vehicles is not functioning. As a result, the two CAVs are unable to cooperate with each other. To avoid reader confusion, we categorize both two situations as Case 1.

\noindent (3) Case 3: CAV-CAV without VIS

This case is widely adopted in cooperative merging papers \cite{rios2016survey}-\cite{chen2023deep}; that is, both on-ramp and mainline vehicles are CAVs, but the vehicle identification process is assumed to magically happen instantaneously when cooperative merging control strategies are designed. In this case, both CAVs start cooperation (Control Mode II) from the beginning of the merging area.

% \begin{table}
% \footnotesize
% \caption{Operation Strategies for the on-ramp CAV in different cases}
% \centering
% \begin{tabular}{p{1.5cm} p{1.3cm} p{1.7cm} p{2.1cm}}   
%     %p paragraph column with text vertically aligned at the top
%     %m paragraph column with text vertically aligned in the middle (requires array package)
%     %b paragraph column with text vertically aligned at the bottom (requires array package)
% \textbf{}    \\ \hline                           
% & \textbf{Before the Start Line} & \multicolumn{2}{c}{\textbf{After the Start Line}}  \\
% \hline
% \textbf{Baseline}                      & \multicolumn{3}{c}{Naturalistic trajectory}   \\
% \hline
% \textbf{Case 1: CAV-THV}               & Naturalistic trajectory        & Naturalistic trajectory (Time to VIS) & Recursive merging control (Control Mode I) \\
% \hline
% \textbf{Case 2: CAV-CAV with VIS}      & Naturalistic trajectory        & Naturalistic trajectory (Time to VIS) & Cooperative merging control (Control Mode II)                                            \\
% \hline
% \textbf{Case 3: CAV-CAV without   VIS} & \multicolumn{3}{c}{\centering Cooperative merging control (Control Mode II)} \\
% \hline                                                                                                       
% \end{tabular}
% \label{table1_operation strategies}
% \end{table}

\begin{table}
\footnotesize
\captionsetup{font=small}
\caption{Operation Strategies for the on-ramp CAV in different cases}
\centering
\begin{tabular}{l}
$\includegraphics[width=3.2in]{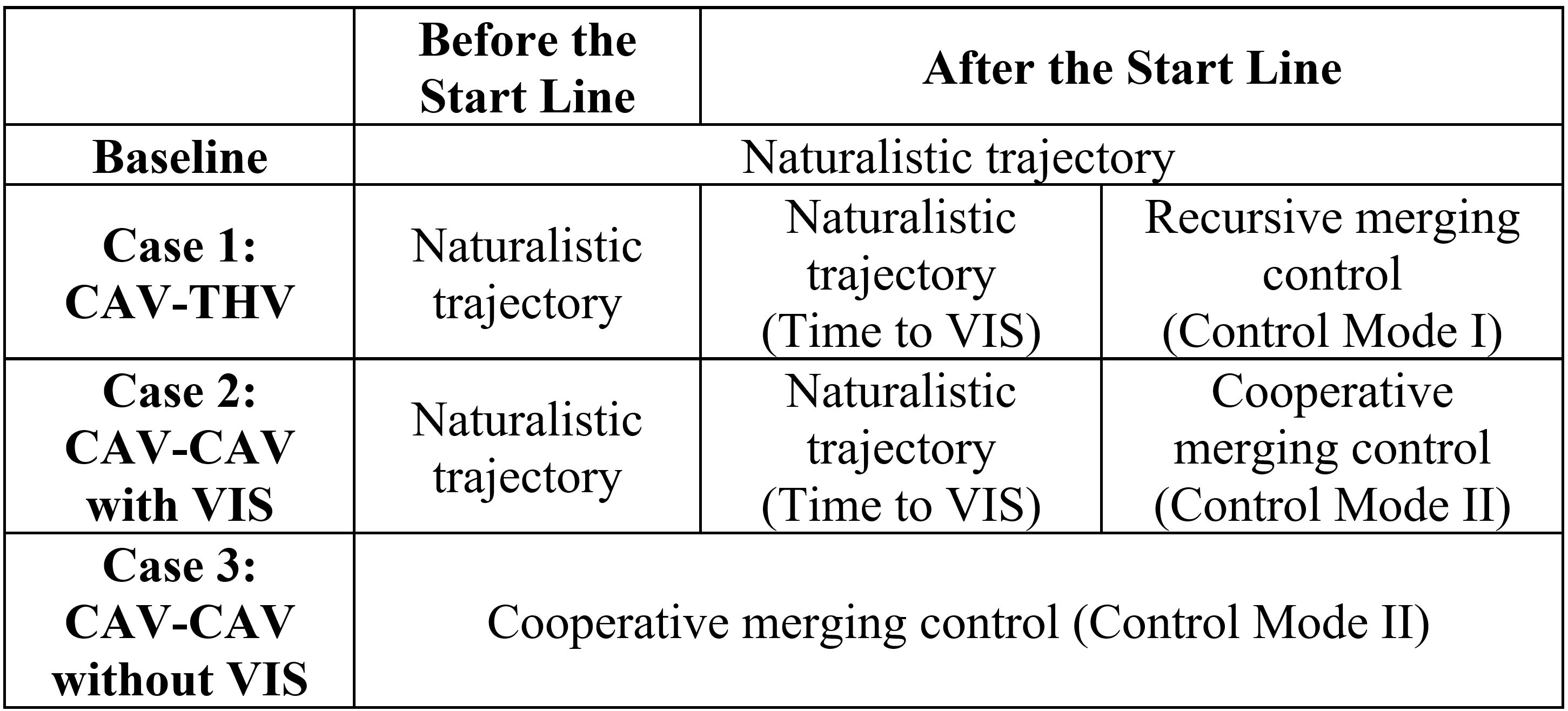}$
\end{tabular}
\label{tab1_operation strategies}
\end{table}

\noindent (4) Evaluation Settings

The abovementioned 3 cases are compared with the baseline case. To achieve fair comparisons, specific settings are applied to all three cases. First, the initial states of vehicles are sampled from the same set-up of the baseline case. The merging positions and times of the baseline pairs are directly used. What’s more, we keep the same merging sequence in the three simulation cases as the baseline. Further, during the period when the on-ramp CAV is employing VIS, we set its trajectory aligned with the baseline case. The practical rationale for adopting this approach can be understood as follows: the on-ramp CAV adheres to its initial driving plan because it is unable to determine which vehicles on the mainline are CAVs. Overall, these setups can fairly evaluate the performance of different cases. The operation strategies for these cases are summarized in Table \ref{tab1_operation strategies}.

\subsection{Connected vehicle identification system (VIS)}

Accurately identifying surrounding connected vehicles is a prerequisite to realize CAV cooperative merging in real-world applications. Specifically, as noted, there are two problems to be solved: 1) which surrounding vehicles are connected vehicles? and 2) which connected vehicles send which V2V messages? Simply relying on GPS positions contained in V2V messages cannot achieve these two tasks because GPS positions have errors varying from $1\sim5$ meters. Therefore, a connected vehicle identification system taking advantage of both radar-measured distances and GPS-measured distances was proposed. 

\begin{figure}
\centering
\includegraphics[width=2in, trim=13cm 3cm 7.8cm 0cm, clip]{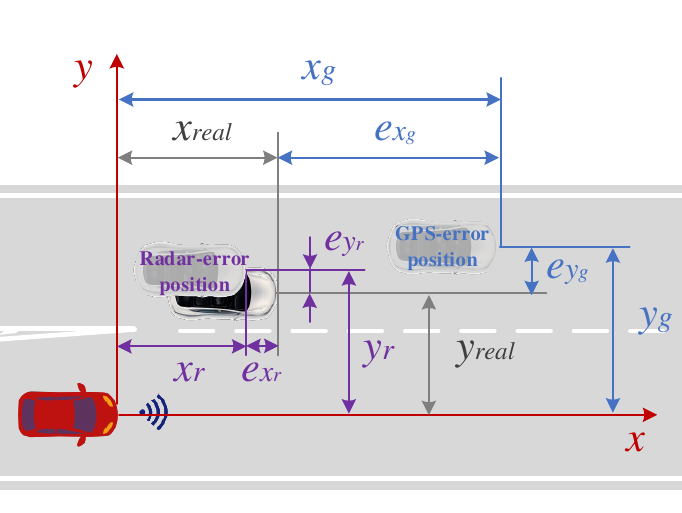}
\captionsetup{justification=justified}
\caption{GPS-measured and radar-measured distance errors.  
}\label{fig5_VIS}
\end{figure}

Consider a coordinate system of an on-ramp CAV with the $x$-axis aligned with its longitudinal direction and the $y$-axis perpendicular to the direction of motion as shown in Fig.~\ref{fig5_VIS}. Denote the real distance between the on-ramp CAV and a target mainline CAV is $(x_{real},y_{real} )$, while the GPS-measured distance and radar-measured distance are $(x_g,y_g )$  and $(x_r,y_r )$, respectively. Assuming that both longitudinal and lateral errors of the GPS $(e_{x_g},e_{y_g})$ and radar $(e_{x_r},e_{y_r})$ follow normal distributions $N\sim(0,\sigma_g=1)$ and $N\sim(0,\sigma_r=1)$ \cite{chen2022connected}. Then, both the longitudinal and lateral difference between GPS and radar $(x_g-x_r,y_g-y_r )$ follow normal distribution $N\sim(0,\sigma=\sqrt{(\sigma_r^2+\sigma_g^2 )}\approx1)$. What’s more, the sum of square of these two differences($x_g-x_r )^2/\sigma^2+(y_g-y_r )^2/\sigma^2$ follows $\chi^2$ distribution. To balance the time and accuracy of identification, for each matching, a threshold $\mathrm{F}^{(-1)} (\chi^2 (2),1-\alpha)$ is needed, which defines the probability $\alpha$ that a specific connected vehicle is within this threshold or not. Based on this we have
\begin{equation} \label{eq:6}
\begin{aligned}
\frac{\left(x_{g}-x_{r}\right)^{2}}{\sigma^{2}}+\frac{\left(y_{g}-y_{r}\right)^{2}}{\sigma^{2}}<\mathrm{F}^{-1}\left(\chi^{2}(2), 1-\alpha\right)
\end{aligned}
\end{equation}
Therefore, identifying a target surrounding connected vehicle means to test if the GPS and radar difference follow chi-square distribution in a certain time.

Due to the limited focus on the VIS process and constraints on space in this paper, we provide a specific identification time applied in the on-ramp area, which is 3.5 seconds. Interested readers can find more details in our previous work \cite{chen2022connected}.

\section{SIMULATION RESULTS AND DISCUSSIONS } \label{section4}
\subsection{Evaluation measures for merging control performance}
Three measures are used to evaluate safety, energy, and comfort performance: vehicle time gap at merging time (merging time gap), mean fuel consumption, and root mean square of accelerations (A-RMS). Merging time gap is deemed a surrogate safety measure, which means the time interval between the time instant the first vehicle last occupied a critical position and the time instant the second vehicle subsequently arrived at the merging position. A Time Gap value of zero indicates a collision at the merging position. A-RMS is used to evaluate the fluctuation of accelerations. The higher the A-RMS value, the more variable the acceleration, resulting in lower comfort for passengers inside the vehicle. Mean Fuel is used to evaluate energy consumption performance during the merging process. The equation of fuel consumption is given \cite{kamal2012model}, which is ruled by vehicle acceleration $u_i$ and speed $v_i$, $i=r$ or $m$ represents an on-ramp vehicle or a mainline vehicle.
\begin{equation} \label{eq:6}
\begin{aligned}
\begin{array}{c}
\dot{f}_{i}=\dot{f}_{c r u, i}(t)+\dot{f}_{a c c, i}(t) \\
\dot{f}_{c r u, i}(t)=\vartheta_{0}+\vartheta_{1} v_{i}(t)+\vartheta_{2} v_{i}{ }^{2}(t)+\vartheta_{3} v_{i}{ }^{3}(t) \\
\dot{f}_{a c c, i}(t)=a_{i}(t)\left(\sigma_{0}+\sigma_{1} v_{i}(t)+\sigma_{2} v_{i}{ }^{2}(t)\right)
\end{array}
\end{aligned}
\end{equation}
\noindent where $ \vartheta_0, \vartheta_1, \vartheta_2, \vartheta_3, \sigma_0, \sigma_1$ and $\sigma_2$ are $0.1569 \mathrm{~mL} / \mathrm{s}$, $0.0245 \mathrm{mL} / \mathrm{m}$, $-7.415 \times 10^{-4} \mathrm{~mL} \cdot \mathrm{s} / \mathrm{m}^{2}$, $5.975 \times 10^{-5} \mathrm{~mL} \cdot \mathrm{s}^{2}/\mathrm{m}^3$, $7.224 \times 10^{-2} \mathrm{mL} \cdot \mathrm{s}/ \mathrm{m}$, $9.681 \times 10^{-2} \mathrm{mL} \cdot \mathrm{s}^2 / \mathrm{m}^2$, and $1.075 \times 10^{-3} \mathrm{mL} \cdot \mathrm{s}^3/ \mathrm{m}^3$. When the vehicle decelerates, i.e., $a_i (t)<0$, the fuel consumption is neglected.

Note that in each case, each measure for each pair of vehicles was first calculated, and then measured values for all pairs were aggregated to obtain an average value. Furthermore, each measure’s value corresponds to time intervals from on-ramp vehicles entering the SL to reaching their merging positions.

\subsection{Merging control results}
Based on the merging control strategies illustrated in Section \ref{section3}, this part reveals how on-ramp CAVs and their corresponding potentially conflicting mainline vehicles perform when on-ramp CAVs utilizing VIS in their merging process. The simulation results corresponding to different merging positions are given in Table II and Table III. 

% \begin{table}
% \footnotesize
% \caption{Operation Strategies for the on-ramp CAV in different cases}
% \centering
%     %p paragraph column with text vertically aligned at the top
%     %m paragraph column with text vertically aligned in the middle (requires array package)
%     %b paragraph column with text vertically aligned at the bottom (requires array package)
% %\begin{tabular}{p{1.5cm} p{1.3cm} p{1.7cm} p{2.1cm}}
% \begin{tabular}{l|ccccc}
% \toprule
% \diagbox [width=5em, trim=1]{sample}{size} & 1 & 2 & 3 & 1 & 1 
% Baseline     & 0.60       & 77.51           & 104.41        & 0.62     & 1.01   \\
% Case   1     & 1.77       & 77.51           & 112.79        & 0.62     & 2.01   \\
% Case   2     & 1.80       & 51.44           & 91.79         & 1.05     & 1.08   \\
% Case   3     & 1.80       & 35.41           & 87.24         & 0.67     & 0.96                                                    
% \end{tabular}
% \label{table2_the 1/3 parts}
% \begin{tablenotes}
%   Note: all results are collected from the starting time to the moment when the on-ramp vehicle reaches the merging point.  
% \end{tablenotes}

% \end{table}

\noindent (1) Comparison of different cases

We first compare the merging safety performance in Table II, which is represented by the merging time gap. Obviously, the baseline case has the smallest average merging time gap since the trajectories we used have high collision risks. In contrast, the vehicles in the other 3 cases are recursively or cooperatively controlled while complying with the safe merging constraints as given in Eqs. (\ref{eq:2})-(\ref{eq:4}), thus keeping safer time gaps from the mainline vehicles. The expected merging time gap, which is set as 1.8 seconds, is realized in Case 2 and Case 3 because CAVs achieve long-term planning via cooperative control. However, in Case 1, the recursive control employed by the on-ramp CAV relies on prediction of the future states of the mainline vehicle in each control step, which cannot be accurate due to the uncertain movements of the mainline vehicle. Therefore, it’s difficult to reach a strictly 1.8-second merging time gap. But still, the average merging time gap of Case 1 is much safer than that of the baseline. Overall, the proposed merging control strategies can effectively mitigate collision risks.

\begin{table}
\footnotesize
\captionsetup{font=small}
\caption{Average results for 100 vehicle pairs when the merging position is in the 1/3 parts of the auxiliary lane}

\centering
\begin{tabular}{l}
$\includegraphics[width=3.2in]{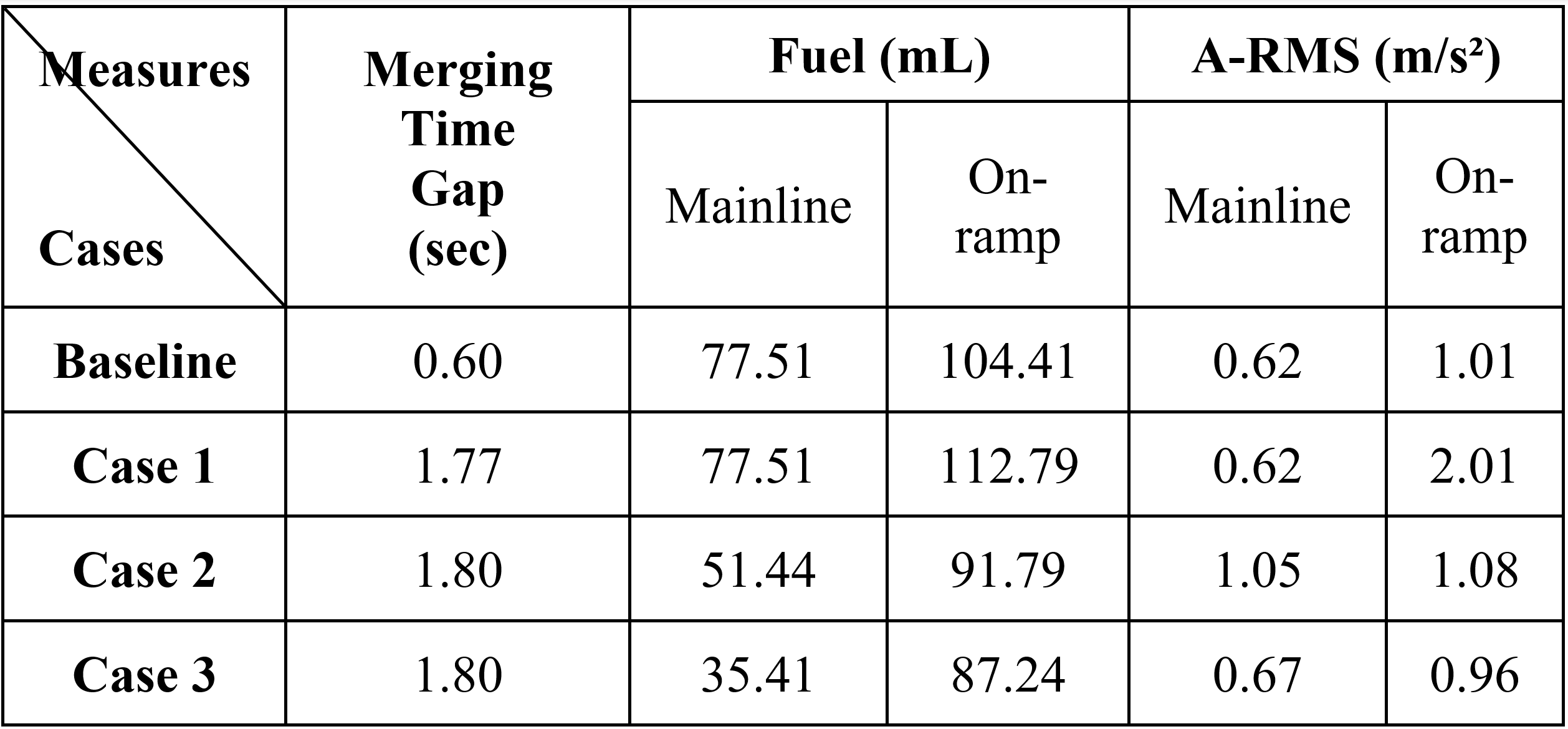}$
\end{tabular}
    \begin{tablenotes}[flushleft] % Start tablenotes environment
      \small
      \item Note: all results are collected from the starting time to the moment when the on-ramp vehicle reaches the merging point. % Define your footnote here
    \end{tablenotes}

\label{tab2_1/3 parts}
\end{table}

The average acceleration root mean square (A-RMS) indicates the comfort performance of different cases, and it also has an impact on fuel consumption. In Case 1, the A-RMS of on-ramp CAV reaches almost twice the A-RMS of the baseline. The reason is that the on-ramp CAV needs to actively create a safe time gap from the mainline THV, whose behavior cannot be accurately predicted. Consequently, the CAV undergoes significant acceleration variations. In comparison to Case 1, the A-RMS of the on-ramp CAV in Case 2 is reduced by nearly half. This improvement is attributed to CAV cooperation, allowing CAVs to plan a longer continuous trajectory without constant adjustments. However, the use of VIS in Case 2 shortens the cooperation time and space, requiring both mainline and on-ramp CAVs to make aggressive acceleration changes to achieve a safe merging gap. As such, the A-RMS in Case 2 remains higher than the baseline's A-RMS. Among all the cases, Case 3 yields the lowest A-RMS of the on-ramp CAV. The A-RMS of the mainline CAV is also lower than that in Case 2. This is because Case 3, without VIS, provides more extended time and space for CAV cooperation. As a result, CAVs have more space to generate a safe merging distance and plan smoother trajectories. 

In terms of fuel consumption, generally, Case 2 and Case 3, which involve CAV cooperation, outperform the other 2 cases. Case 3 has the best fuel consumption performance due to its long-term trajectory planning. Case 1 has the highest fuel consumption among all cases. This is because the on-ramp CAV in Case 1 needs to adjust its trajectory recursively according to the uncertain movements of the mainline vehicle, causing more acceleration changes and leading to more fuel consumption. Besides, vehicles achieve a larger merging time gap compared to the baseline case, which means that the on-ramp CAV improves merging safety at the cost of increased energy loss and decreased comfort. It is noted that the A-RMS and fuel consumption differences between Case 2 and Case 3 are statistically significant at the 95$^{\text{th}}$ percentile confidence level. 

It's noteworthy that, the higher fuel consumption values for on-ramp vehicles compared to mainline vehicles can be attributed to our calculation method. That is, we calculate fuel consumption from the starting time to the moment when the on-ramp vehicle reaches the merging point. Since 82\% of the generated trajectories involve the on-ramp vehicle leading the mainline vehicle at the merging position, it is reasonable that on-ramp vehicles travel longer distances and have more fuel consumption.

\begin{figure}
\centering
\includegraphics[width=3.4in]{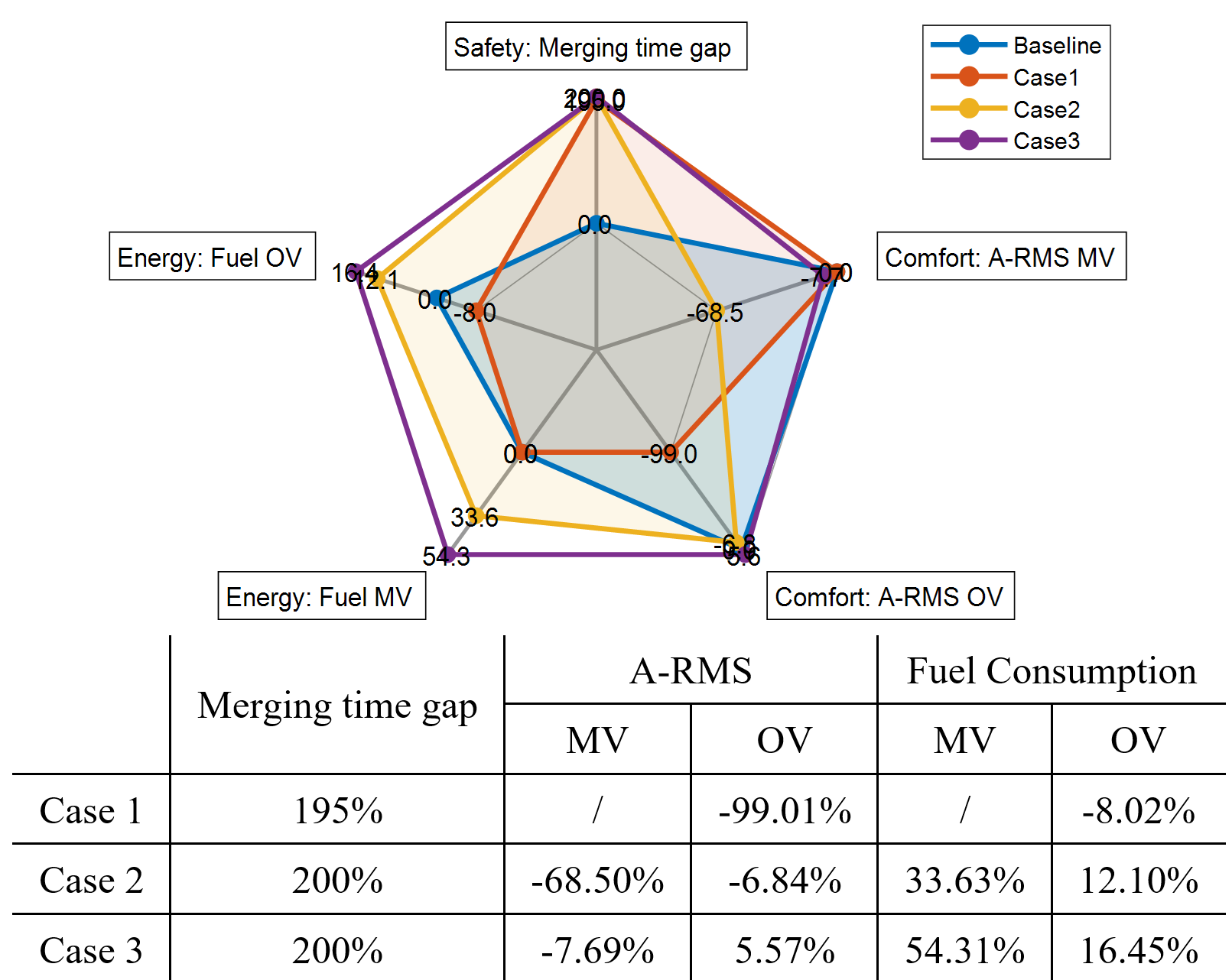}
\captionsetup{justification=justified}
\caption{Comparison of merging performance when the merging position is in the 1/3 parts of the auxiliary lane. Each number represents the improvement rate (\%) of each measure in each case. MV means the mainline vehicle, OV means the on-ramp CAV.  
}\label{fig6_radarfigure}
\end{figure}

The percentage of performance improvement of each measure is presented in Fig. \ref{fig6_radarfigure}. This figure intuitively illustrates the performance of merging safety, energy efficiency, and comfort through the region occupied by each case. In general, CAV control enhances merging safety, and CAV cooperation goes a step further by improving energy efficiency and comfort while ensuring safety. VIS may introduce a delay in CAV cooperation time, causing an uncomfortable driving experience given a limited cooperation space and time. However, it is essential as it makes CAV cooperation possible.

\noindent (2) Impact of merging positions

The merging performance under the impact of the merging position can be evaluated by comparing Table \ref{tab2_1/3 parts} and Table \ref{tab3_2/3 parts}. Basically, the differences in measures across different cases in Table \ref{tab3_2/3 parts} adhere to similar patterns as those observed in Table II. However, there are still some differences worth analyzing. The most significant change is the increase in fuel consumption in Table III, which is caused by the longer travel distance before merging. Besides, since a longer travel distance provides vehicles with a longer space and time to adjust their states, the average merging time gap of Case 1 increases from 1.77 sec to 1.78 sec, indicating a safer merging process. What’s more, in terms of average fuel consumption of on-ramp vehicles, values in Table \ref{tab3_2/3 parts}-Cases 2 (97.37 mL) and 3 (91.81 mL) are lower than the value in Table \ref{tab2_1/3 parts}-Baseline (104.41 mL), despite the on-ramp vehicles in Table III travel longer distances. This once again highlights the advantages of cooperative merging.

\begin{table}
\footnotesize
\captionsetup{font=small}
\caption{Average results for 100 vehicle pairs when the merging position is in the 2/3 parts of the auxiliary lane}
\centering
\begin{tabular}{l}
$\includegraphics[width=3.2in]{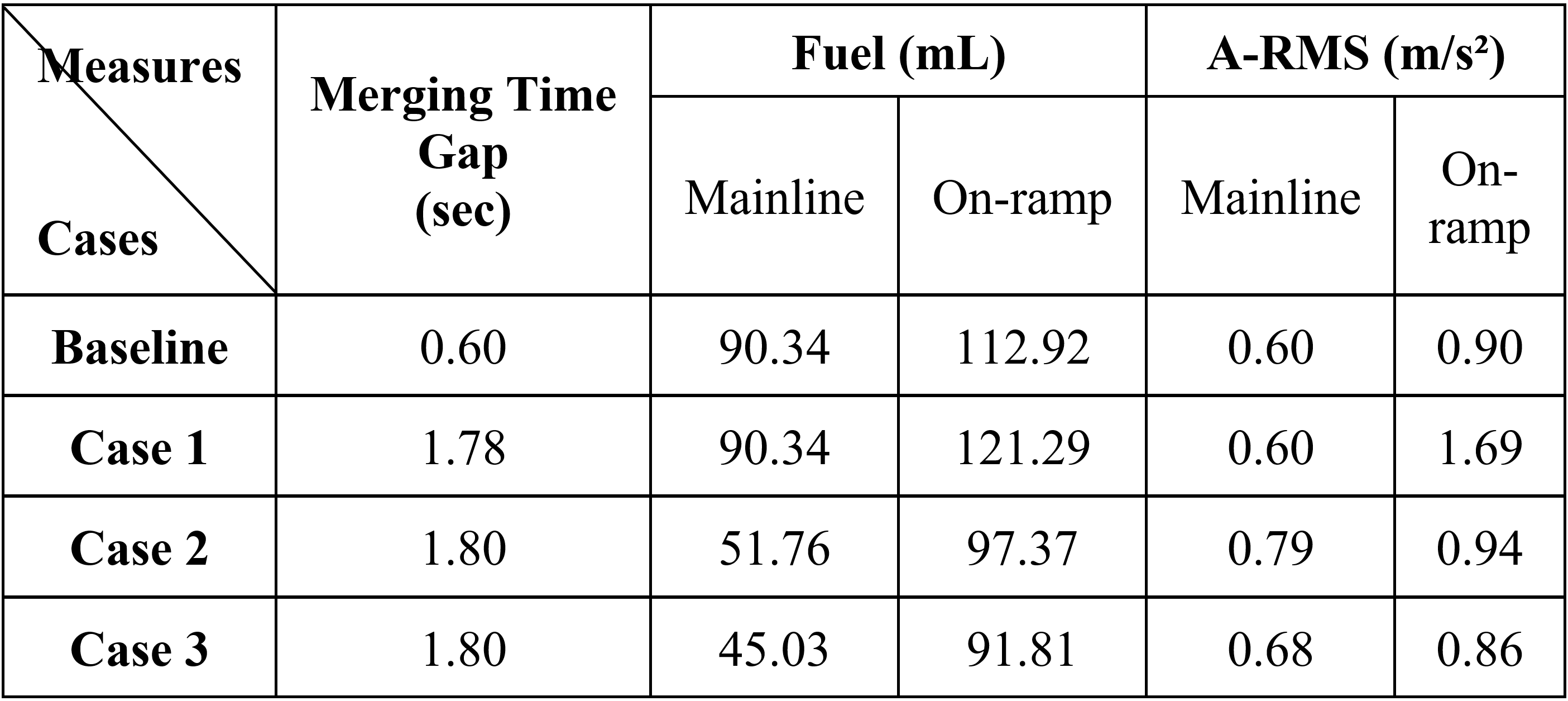}$
\end{tabular}
    \begin{tablenotes} [flushleft]% Start tablenotes environment
      \small
      \item Note: all results are collected from the starting time to the moment when the on-ramp vehicle reaches the merging point. % Define your footnote here
    \end{tablenotes}
\label{tab3_2/3 parts}
\end{table}

The percentage of A-RMS and fuel consumption reduction is further quantified and compared in Fig. 7. This figure reveals that with the increase in travel distance, on-ramp CAVs have more ample space and time for adjusting their motion, allowing for an enhancement in merging safety, coupled with a simultaneous consideration enhancement in energy efficiency and comfort.

\begin{figure}
\centering
\includegraphics[width=3.4in]{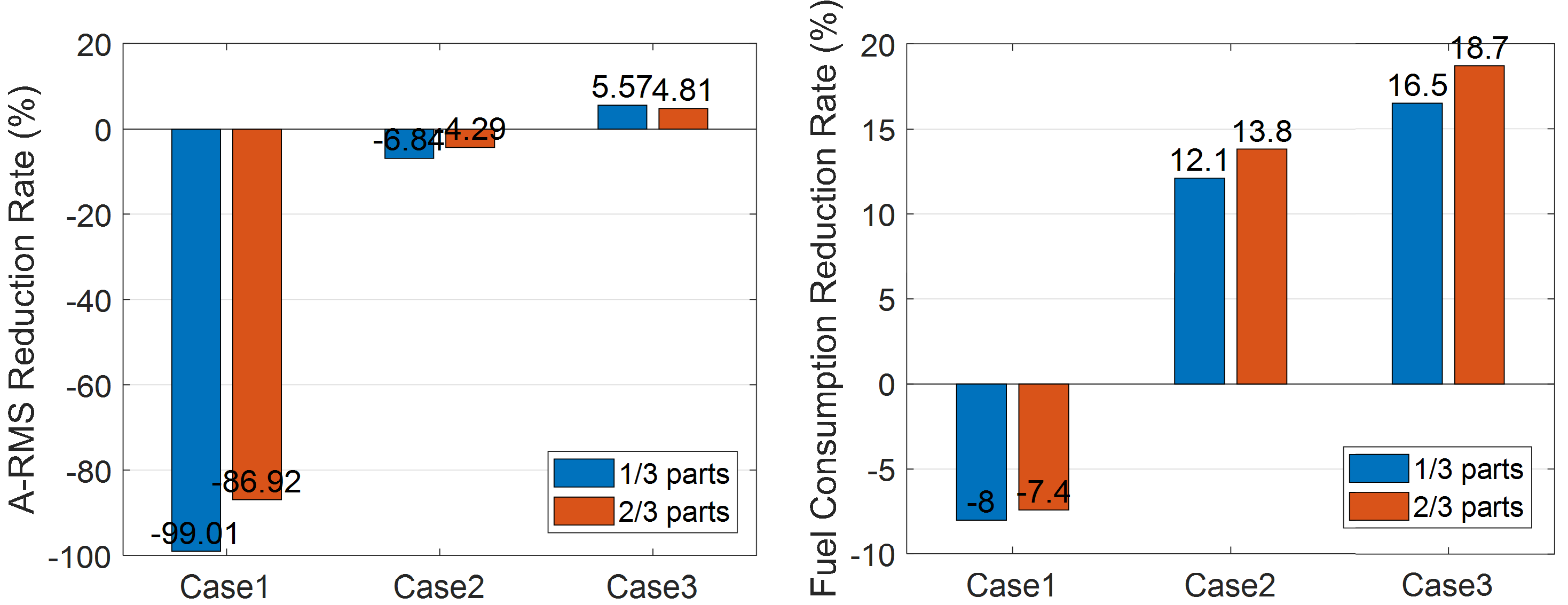}
\captionsetup{justification=justified}
\caption{Comparison of A-RMS and fuel consumption reduction rate of the on-ramp CAV when the merging positions are in different merging areas.  
}\label{fig7_comparison}
\end{figure}

\subsection{Discussions}

\noindent (1) Impact of vehicle identification process

Simulation results provide insights into how the vehicle identification process impacts CAV merging performance. An average identification time of 3.5 seconds is observed in on-ramp merging areas. Given the high-speed nature of vehicles in highway scenarios and the constrained space on the on-ramp, this 3.5-second period is critical for CAVs to make necessary adjustments and enhance driving performance. Based on the identification time, the challenge posed by VIS to merging control, when the line of sight is limited, and the auxiliary lane is short, is highlighted in the comparison of A-RMS. That is, VIS time shortens merging cooperation time and space, which may lead CAVs to take aggressive actions in a limited space and time to generate safe time gap between the mainline vehicle, thus reducing the benefits of CAV cooperation.

However, the advantage of VIS is evident when comparing Case 1 and Case 2. In Case 1, there is a situation where both mainline and on-ramp vehicles are CAVs but the VIS of both vehicles is not functioning. Consequently, the on-ramp CAV can only adopt recursive merging control, resulting in inferior merging performance compared to Case 2, which involves CAV cooperation, and even worse than the baseline case.

\noindent (2) Benefits of CAV traffic and CAV cooperation

The presence of CAVs effectively improves merging safety, as seen in the comparison of merging time gaps in the baseline case and those in the three cases where CAVs exist. The benefits of CAV cooperation are demonstrated through comparisons between different cases, particularly in Case 3, which has a longer CAV cooperation range and yields the best merging performance of all measures.

\noindent (3) Insights into infrastructure-side improvements

The simulation results shed light on potential infrastructure improvements for improving merging safety and ecology. Analyzing the improvement rate depicted in Figs.~\ref{fig6_radarfigure} and \ref{fig7_comparison} reveals that more significant enhancements in merging performance can be achieved when vehicles travel a greater distance, affording them ample space and time for adjustments. Additionally, the superiority of Case 3 over Case 2 underscores the advantages of expanding the cooperation range. These findings may lead people to think that a longer auxiliary lane design contributes to realizing superior merging performance. However, refining existing road structures requires high cost; plus, even with an additional hundred meters of auxiliary lane length, as evaluated in this paper, the improvement rate of each measure is limited. Therefore, extending the length of the on-ramp or the auxiliary lane is not a cost-effective option. 

Considering the impracticality of unlimited extension of auxiliary lanes in real-world scenarios, there are two alternative strategies to enhance safety in merging areas. One approach involves the removal of visual obstacles near the merging area, allowing CAVs to extend their detection range with an unobstructed line of sight. The second strategy entails the installation of Roadside Units (RSUs) in on-ramp merging areas, equipped with detection and communication devices. This enables CAVs to access more information through Vehicle-to-Infrastructure (V2I) communication, significantly expanding the detection and communication range compared to relying solely on V2V communication. Moreover, RSUs can detect THVs and encode their information into virtual Basic Safety Messages (BSMs) for broadcast. As a result, CAVs gain advanced knowledge of the precise position of each CAV and THV, along with the unique ID of each CAV for establishing cooperation connections. This renders the detrimental impact of VIS time negligible, with the advantages of CAV cooperation prevailing.

\section{CONCLUSIONS}  \label{section5}

Aiming at bridging the gap between the theoretical investigation and real-world application of cooperative on-ramp merging control of CAVs, this paper for the first time introduced the connected vehicle identification system (VIS) into the CAV merging control task. The importance of VIS was illustrated, and the methodology of vehicle identification was elaborated. Then, real-world naturalistic merging trajectory dataset was processed to generate hazardous merging scenarios. To improve merging safety in mixed traffic, different merging operation strategies for CAVs in different mixed-traffic cases were proposed considering VIS process. Simulation results reveal that CAV control contributes to enhanced merging safety, with CAV cooperation further improving energy efficiency and comfort over the baseline (that does not consider CAVs). The introduction of VIS delays the implementation time of cooperation and reduces the space of cooperative merging. This results in on-ramp CAVs sacrificing comfort and fuel consumption to achieve a safe merging distance. Additionally, an increase in travel distance affords CAVs more space and time to adjust their motion, thereby enhancing merging safety and considering improvements in energy efficiency and comfort. As noted in the discussion section, a Roadside Unit (RSU) at the freeway merge area where line of sight is limited and the auxiliary lane is short, would enable CAVs to gain additional information through V2I communication, expanding ranges for detecting THVs and identifying CAVs significantly. Consequently, it would contribute to further enhancing merging safety and addressing ecological considerations.

In terms of future research, our proposed control strategies involve merging time gap enlargement to ensure merging safety, which might be not possible in dense traffic where surrounding vehicles have tight distances between the controlled vehicle pairs. This trickier scenario will be studied in our future research. Additionally, our research will be extended to multi-lane scenarios with multi-vehicle interactions, enhancing merging flexibility and safety.

%%%%%%%%%%%%%%%%%%%%%%%%%% References
\bibliographystyle{IEEEtran}
%\bibliography{IEEEabrv,refs}

\begin{thebibliography}{10}
\providecommand{\url}[1]{#1}
\csname url@samestyle\endcsname
\providecommand{\newblock}{\relax}
\providecommand{\bibinfo}[2]{#2}
\providecommand{\BIBentrySTDinterwordspacing}{\spaceskip=0pt\relax}
\providecommand{\BIBentryALTinterwordstretchfactor}{4}
\providecommand{\BIBentryALTinterwordspacing}{\spaceskip=\fontdimen2\font plus
\BIBentryALTinterwordstretchfactor\fontdimen3\font minus \fontdimen4\font\relax}
\providecommand{\BIBforeignlanguage}[2]{{%
\expandafter\ifx\csname l@#1\endcsname\relax
\typeout{** WARNING: IEEEtran.bst: No hyphenation pattern has been}%
\typeout{** loaded for the language `#1'. Using the pattern for}%
\typeout{** the default language instead.}%
\else
\language=\csname l@#1\endcsname
\fi
#2}}
\providecommand{\BIBdecl}{\relax}
\BIBdecl

\bibitem{guanetti2018control}
J.~Guanetti, Y.~Kim, and F.~Borrelli, ``Control of connected and automated vehicles: State of the art and future challenges,'' \emph{Annual reviews in control}, vol.~45, pp. 18--40, 2018.

\bibitem{liu2023safety}
H.~Liu, W.~Zhuang, G.~Yin, Z.~Li, and D.~Cao, ``Safety-critical and flexible cooperative on-ramp merging control of connected and automated vehicles in mixed traffic,'' \emph{IEEE Transactions on Intelligent Transportation Systems}, vol.~24, no.~3, pp. 2920--2934, 2023.

\bibitem{rios2016survey}
J.~Rios-Torres and A.~A. Malikopoulos, ``A survey on the coordination of connected and automated vehicles at intersections and merging at highway on-ramps,'' \emph{IEEE Transactions on Intelligent Transportation Systems}, vol.~18, no.~5, pp. 1066--1077, 2016.

\bibitem{ding2019rule}
J.~Ding, L.~Li, H.~Peng, and Y.~Zhang, ``A rule-based cooperative merging strategy for connected and automated vehicles,'' \emph{IEEE Transactions on Intelligent Transportation Systems}, vol.~21, no.~8, pp. 3436--3446, 2019.

\bibitem{pei2019cooperative}
H.~Pei, S.~Feng, Y.~Zhang, and D.~Yao, ``A cooperative driving strategy for merging at on-ramps based on dynamic programming,'' \emph{IEEE Transactions on Vehicular Technology}, vol.~68, no.~12, pp. 11\,646--11\,656, 2019.

\bibitem{jing2019cooperative}
S.~Jing, F.~Hui, X.~Zhao, J.~Rios-Torres, and A.~J. Khattak, ``Cooperative game approach to optimal merging sequence and on-ramp merging control of connected and automated vehicles,'' \emph{IEEE Transactions on Intelligent Transportation Systems}, vol.~20, no.~11, pp. 4234--4244, 2019.

\bibitem{shi2023cooperative}
J.~Shi, K.~Li, C.~Chen, W.~Kong, and Y.~Luo, ``Cooperative merging strategy in mixed traffic based on optimal final-state phase diagram with flexible highway merging points,'' \emph{IEEE Transactions on Intelligent Transportation Systems}, 2023.

\bibitem{xiao2021decentralized}
W.~Xiao and C.~G. Cassandras, ``Decentralized optimal merging control for connected and automated vehicles with safety constraint guarantees,'' \emph{Automatica}, vol. 123, p. 109333, 2021.

\bibitem{liu2018strategy}
H.~Liu, W.~Zhuang, G.~Yin, Z.~Tang, and L.~Xu, ``Strategy for heterogeneous vehicular platoons merging in automated highway system,'' in \emph{2018 chinese control and decision conference (ccdc)}.\hskip 1em plus 0.5em minus 0.4em\relax IEEE, 2018, pp. 2736--2740.

\bibitem{liu2021decentralized}
H.~Liu, W.~Zhuang, G.~Yin, R.~Li, C.~Liu, and S.~Zhou, ``Decentralized on-ramp merging control of connected and automated vehicles in the mixed traffic using control barrier functions,'' in \emph{2021 IEEE International Intelligent Transportation Systems Conference (ITSC)}.\hskip 1em plus 0.5em minus 0.4em\relax IEEE, 2021, pp. 1125--1131.

\bibitem{chen2023deep}
D.~Chen, M.~R. Hajidavalloo, Z.~Li, K.~Chen, Y.~Wang, L.~Jiang, and Y.~Wang, ``Deep multi-agent reinforcement learning for highway on-ramp merging in mixed traffic,'' \emph{IEEE Transactions on Intelligent Transportation Systems}, 2023.

\bibitem{rychlicki2020analysis}
M.~Rychlicki, Z.~Kasprzyk, and A.~Rosi{\'n}ski, ``Analysis of accuracy and reliability of different types of gps receivers,'' \emph{Sensors}, vol.~20, no.~22, p. 6498, 2020.

\bibitem{chen2022connected}
Z.~Chen and B.~B. Park, ``Connected preceding vehicle identification for enabling cooperative automated driving in mixed traffic,'' \emph{Journal of transportation engineering, Part A: Systems}, vol. 148, no.~5, p. 04022013, 2022.

\bibitem{wang2022mobility}
Z.~Wang, R.~Gupta, K.~Han, H.~Wang, A.~Ganlath, N.~Ammar, and P.~Tiwari, ``Mobility digital twin: Concept, architecture, case study, and future challenges,'' \emph{IEEE Internet of Things Journal}, vol.~9, no.~18, pp. 17\,452--17\,467, 2022.

\bibitem{chen2024safety}
P.~Chen, H.~Ni, L.~Wang, G.~Yu, and J.~Sun, ``Safety performance evaluation of freeway merging areas under autonomous vehicles environment using a co-simulation platform,'' \emph{Accident Analysis \& Prevention}, vol. 199, p. 107530, 2024.

\bibitem{tian2018performance}
D.~Tian, G.~Wu, K.~Boriboonsomsin, and M.~J. Barth, ``Performance measurement evaluation framework and co-benefit$\backslash$/tradeoff analysis for connected and automated vehicles (cav) applications: A survey,'' \emph{IEEE Intelligent Transportation Systems Magazine}, vol.~10, no.~3, pp. 110--122, 2018.

\bibitem{moers2022exid}
T.~Moers, L.~Vater, R.~Krajewski, J.~Bock, A.~Zlocki, and L.~Eckstein, ``The exid dataset: A real-world trajectory dataset of highly interactive highway scenarios in germany,'' in \emph{2022 IEEE Intelligent Vehicles Symposium (IV)}.\hskip 1em plus 0.5em minus 0.4em\relax IEEE, 2022, pp. 958--964.

\bibitem{zhou2018optimal}
Y.~Zhou, M.~E. Cholette, A.~Bhaskar, and E.~Chung, ``Optimal vehicle trajectory planning with control constraints and recursive implementation for automated on-ramp merging,'' \emph{IEEE Transactions on Intelligent Transportation Systems}, vol.~20, no.~9, pp. 3409--3420, 2018.

\bibitem{vogel2003comparison}
K.~Vogel, ``A comparison of headway and time to collision as safety indicators,'' \emph{Accident analysis \& prevention}, vol.~35, no.~3, pp. 427--433, 2003.

\bibitem{mu2023does}
Z.~Mu, F.~Jahedinia, and B.~B. Park, ``Does the intelligent driver model adequately represent human drivers?'' in \emph{VEHITS}, 2023, pp. 113--121.

\bibitem{kamal2012model}
M.~A.~S. Kamal, M.~Mukai, J.~Murata, and T.~Kawabe, ``Model predictive control of vehicles on urban roads for improved fuel economy,'' \emph{IEEE Transactions on control systems technology}, vol.~21, no.~3, pp. 831--841, 2012.

\end{thebibliography}

\end{document}